\documentclass[a4paper]{article}
\usepackage[latin1]{inputenc}
\usepackage{latexsym}
\usepackage{euscript}
\usepackage{amssymb,amsmath,amsfonts,amstext,amsxtra}
\usepackage{amscd,amsbsy,amsopn}

\newcommand{\deltabar}{\,\,{\bar{}\hspace{1pt}\! \!\delta }}

\begin{document}

\title{Test particles behavior in the framework of a lagrangian geometric 
theory with propagating torsion}
\author{G. Aprea, G. Montani, R. Ruffini\\
\small{ICRA-International Center for Relativistic Astrophysics,}\\
\footnotesize{Dipartimento di Fisica, Universit\`a degli Studi di Roma ``La 
Sapienza",}\\
\footnotesize{Piazzale Aldo Moro 5, 00185 Roma, Italy.}}

\date{April 10, 2003}

\maketitle

\begin{abstract}
Working in the lagrangian framework, we develop a geometric theory in vacuum 
with propagating torsion; the antisymmetric and trace parts of the torsion 
tensor, considered as derived from local potential fields, are taken and, 
using the minimal action principle, their field equations are calculated. 
Actually these will show themselves to be just equations for propagating 
waves giving torsion a behavior similar to that of metric which, as 
known, propagates through gravitational waves.
Then we establish a principle of minimal substitution to derive test particles 
equation of motion, obtaining, as result, that they move along 
autoparallels. We then calculate the analogous of the geodesic deviation for 
these trajectories and analyze their behavior in the nonrelativistic limit, 
showing that the torsion trace potential $\phi$ has a phenomenology which 
is indistinguishable from that of the gravitational newtonian field; in 
this way we also give a reason for why there have never been evidence for 
it. 
\end{abstract}

\begin{flushleft}
\end{flushleft}

\small{

\section{Introduction}

As well known, in non flat spaces the concept of parallel transport of 
vector fields needs the introduction of a connection which is also needed 
to define the covariant derivative. In fact, by means of a connection, 
we can define the equation of parallel transport in the following way: on 
a manifold $M$, given a curve $\gamma(t)$ passing for a point $P\in M$, 
the parallel transported vector field of the vector field $V^{\alpha}(P)$ 
along $\gamma(t)$ is the solution of:
\begin{align}
\frac{dV^{\alpha}}{dt}=
-C_{\mu\nu}^{\phantom{\mu\nu}\alpha}V^{\nu}\dot{\gamma}^{\mu}.
\end{align}
In regard to the covariant derivative of a vector field 
$V^{\alpha}(x)$, instead, it is defined in this way:
\begin{align}
D_{\beta}V^{\alpha}=\partial_{\beta}V^{\alpha}+
C_{\beta\gamma}^{\phantom{\beta\gamma}\alpha}V^{\gamma}
\end{align}
The quantities $C_{\mu\nu}^{\phantom{\mu\nu}\alpha}$ are just the 
coefficients of the affine connection and a space, endowed with them, is 
called an affine space and usually indicated by the symbol $L_4$; in 
general these coefficients are non tensorial quantities but 
that is not true for their antisymmetric part called torsion:
\begin{align}                                                         
S_{\alpha\beta}^{\phantom{\alpha\beta}}=
C_{[\alpha\beta]}^{\phantom{[\alpha\beta]}\alpha},
\end{align}
which indeed transforms like a tensor. Just because of this property, the 
presence of torsion in space-time denies the principle of equivalence of 
its importance; here we are not referring to the equivalence between 
inertial and gravitational mass, which is preserved since the theory remains 
geometric, but to the formulation of the equivalence principle \cite{msnr} 
according to which, once defined an inertial
\footnote{In an inertial frame, a body at rest remain in such a state.} 
frame in a point, there the laws of physics are the same as those of special 
relativity. In the case of presence of torsion, the latter, being a tensor, 
can't be set to zero by a convenient choice of the coordinates so, since we 
expect torsion to be source of some force, it is not possible to define an 
inertial frame in any point which is a necessary condition for the 
applicability of the principle.\\

In order to calculate lengths and angles, we also need to introduce 
in our affine space $L_4$ a metric and that can be done defining the 
square modulus of a vector $V^{\alpha}$ as:
\begin{align}
\|V\|^2=g_{\alpha\beta}V^{\alpha}V^{\beta};
\end{align}
here $g_{\alpha\beta}$ is just the symmetric metric tensor. 
With the help of the tensor of nonmetricity:
\begin{align}
Q_{\alpha\beta\gamma}\equiv-D_{\alpha}g_{\beta\gamma},
\end{align}
it is possible to establish a relation between the connection 
coefficients and torsion, metric, and nonmetricity:
\begin{align}                                                          
                                                       \label{e:1_1}
\begin{split}
&C_{\alpha\beta\gamma}=
\frac{1}{2}\left[
\partial_{\alpha}g_{\beta\gamma}-
\partial_{\gamma}g_{\alpha\beta}+
\partial_{\beta}g_{\alpha\gamma}\right]+
\left[S_{\alpha\beta\gamma}-
S_{\beta\gamma\alpha}+
S_{\gamma\alpha\beta}\right]+\\&+
\frac{1}{2}\left[
Q_{\alpha\beta\gamma}-
Q_{\gamma\beta\alpha}+
Q_{\beta\gamma\alpha}\right]\equiv
\Gamma_{\alpha\beta\gamma}+
K_{\alpha\beta\gamma}+
\frac{1}{2}\left[
Q_{\alpha\beta\gamma}-
Q_{\gamma\beta\alpha}+
Q_{\beta\gamma\alpha}\right].  
\end{split}
\end{align}
The new quantities we have introduced here are the Christoffel symbols
\footnote{Although the tensorial formalism used, they are non tensorial 
quantities} 
$\Gamma_{\beta\alpha\gamma}$ and the contortion tensor 
$K_{\alpha\beta\gamma}$;
the former are symmetric in the first two indices while the latter is 
antisymmetric in the last two. In the rest of this paper, in order to 
preserve lengths and angles under parallel displacement, we will assume 
the metric postulate according to which nonmetricity is vanishing; such a 
space is indicated by the symbol $U_4$:
$$
\begin{CD}
\fbox{\scriptsize{General linear space $L_4$}}
@>{Q_{\alpha\beta\gamma}}>>
\fbox{\scriptsize{Einstein-Cartan space $U_4$}}
@>{S_{\alpha\beta\gamma}}>>
\fbox{\scriptsize{Riemannian space $V_4$}}
\end{CD}
$$
Completely neglected in the first formulation of the theory of General 
Relativity (GR) by Einstein, the introduction of torsion was later taken 
into consideration by Einstein himself \cite{nstn}, Eddington \cite{ddngtn}, 
Schr\"odinger \cite{schrdngr} and principally Cartan \cite{crtn1}\cite{crtn2}
\cite{crtn3}, who had the idea of a theory in which torsion was 
connected with intrinsic angular momentum. Later this idea was shelved 
and only since the fifties the possibility of introducing torsion into GR 
had been revalued. Utiyama \cite{tym}, Kibble \cite{kbbl} and Sciama 
\cite{scm1}\cite{scm2}, inspired by the work of Yang and Mills \cite{yng} 
on gauge theories, formulated a theory of gravitation as a gauge theory in 
which the presence of torsion was necessary; then by the mid seventies 
Hehl et al. \cite{hhl1} managed to set up a Poincar\'e gauge theory of 
gravitation, the $U_4$ theory, with torsion corresponding to the 
rotation gauge potential. They assumed the geometric lagrangian density 
was the curvature scalar and that matter could be taken into account simply 
through the minimal coupling rules: $\eta_{\mu\nu} \to g_{\mu\nu}$, 
$\partial_{\mu} \to D_{\mu}$; in this theory intrinsic angular momentum 
creates torsion which can't propagate through empty space; in brief, this 
is due to the fact that the field equations that relate torsion and spin are 
algebraic and don't involve derivatives, allowing the substitution of 
torsion in the matter Lagrangian leading to an effective contact interaction. 
Since, in the first instance, it is reasonable to expect torsion to behave 
as any other interaction field, i. e. propagating into vacuum, this aspect 
of $U_4$ theory is unsatisfactory and in this paper a theory to overcome 
this problem is discussed. In the next section some assumptions are made 
about the form of the torsion tensor which are necessary to 
obtain a propagating torsion without changing the form of the action, 
namely the completely antisymmetric and trace part of the torsion 
tensor are considered derived from two local torsion potential. Then, 
in section \ref{s:3}, by the principle of least action we determine the 
field equations for these potentials which indeed reveal themselves to 
be wave equations. In the second part (sections \ref{s:4} 
and \ref{s:5}) we discuss the problem of determining the 
equation of motion of test particles and establish a principle of minimal 
substitution which leads us to say that test particles move along 
autoparallels; finally we calculate the nonrelativistic limit of these 
trajectories and of the tidal effects and show that the torsion trace potential 
$\phi$ enters in all the equations in the same way as, in this limit, the 
gravitational potential does. Concluding remarks follow.

\section{The form of the torsion tensor}                     \label{s:2}

As we have seen, torsion is a three indices tensor which is antisymmetric 
in the first two; according to the rules of group theory it can be 
decomposed 
in a completely antisymmetric part, a trace part and a third part with 
no special symmetry properties \cite{hhl2}. In our analysis we consider 
only the 
first two terms and, in addition, we assume that they can be derived from 
the exterior derivative of two potentials, in this way:
\begin{align}
S_{[\mu\nu\lambda]} &\to 
B_{\mu\nu\lambda}\equiv
\partial_{[\mu}A_{\nu\lambda]}=
\overset{\{\}}{D}_{[\mu}A_{\nu\lambda]}        \label{e:2_1}\\
Tr(S_{\mu\nu\lambda}) &\to 
\frac{1}{3}
((\partial_\mu\phi)g_{\nu\lambda}-
(\partial_\nu\phi)g_{\mu\lambda})  
                                               \label{e:2_2}
\end{align}
$A_{\mu\nu}(x)$ is an antisymmetric tensor while $\phi(x)$ is a scalar.\\
These potentials play a role analogous to that of metric in the symmetric 
part of the connection:
\begin{align}
\Gamma_{\alpha\beta\gamma}=
\frac{1}{2}\left[
\partial_{\alpha}g_{\beta\gamma}-
\partial_{\gamma}g_{\alpha\beta}+
\partial_{\beta}g_{\alpha\gamma}\right].
\end{align}
By virtue of (\ref{e:1_1}), we can also write the expression for the connection 
coefficients and contortion:
\begin{align}
\begin{split}
K_{\mu\nu\lambda} &\equiv 
S_{\mu\nu\lambda}-S_{\nu\lambda\mu}+S_{\lambda\mu\nu}= \\ 
&=\partial_{[\mu}A_{\nu\lambda]}+
\frac{2}{3}((\partial_\lambda\phi)g_{\mu\nu}-
(\partial_\nu\phi)g_{\mu\lambda})  
                                                   \label{e:2_3}
\end{split}\\
\begin{split}
C_{\mu\nu\lambda} &=\Gamma_{\mu\nu\lambda}+K_{\mu\nu\lambda}=\\ 
&=\Gamma_{\mu\nu\lambda}+\partial_{[\mu}A_{\nu\lambda]}+
\frac{2}{3}((\partial_\lambda\phi)g_{\mu\nu}-
(\partial_\nu\phi)g_{\mu\lambda})  
                                                   \label{e:2_4}
\end{split}
\end{align}
The introduction of torsion potentials for the antisymmetric part of 
torsion is already present in the literature
\footnote{See, for example, \cite{hmmnd}} and has its main motivation 
just in obtaining a torsion propagating into vacuum. As far as the 
expression (\ref{e:2_2}) for the trace part is concerned, it is worth 
noting that the same expression is present in \cite{hjmn} but in a different 
contest; in fact, in Hojman et al's article it was introduced to get a 
coupling of torsion to electromagnetic field which didn't break gauge 
symmetry. Really there is another way available to obtain propagation; it 
consists, in analogy to Yang Mills theory, in introducing square terms in 
curvature and torsion in the Einstein-Hilbert action. This approach is 
discussed, among others, in \cite{szgn1},\cite{nvll},\cite{szgn2},
\cite{tsytln},\cite{khfss},\cite{blgjvc}; here we make the different 
choice of using torsion potentials which, we believe, has these 
advantages: 1) we can preserve the simplicity of the Einstein-Hilbert 
action with the minimal substitution $\Gamma_{\alpha\beta\gamma} \to 
G_{\alpha\beta\gamma} + K_{\alpha\beta\gamma}$; 2) we have put both 
riemannian connection and torsion on the same level since as the former 
is derived from metric, the latter is derived from potentials; 3) in the 
limit of small and slow varying $\phi$ the action (\ref{e:3_0}) is equivalent
to the low energy limit of string theory lagrangian, as already mentioned 
in \cite{hmmnd}(and reference therein), suggesting torsion potentials to be 
a necessary ingredient in more general theory.

\section[Field equations]{Field equations}                     \label{s:3}

According to what we have said in the previous sections our Lagrangian 
density is of the Hilbert-Einstein form
\footnote
{
Here $k$ is a constant related to Newton's gravitational constant 
$G=6.670\times 10^{-8}cm^3g^{-1}s^{-2}$ via:
\begin{align*}
\frac{1}{k}=\frac{c^3}{8\pi G}
\end{align*}
}:
\begin{flalign}                                                  
\begin{split}
\mathcal{A} &= 
-\frac{1}{2k}\int dx \ \sqrt{-g} \ R(x)\equiv\\
&\equiv
-\frac{1}{2k}\int dx \ \sqrt{-g} \ 
g^{\beta\gamma}\delta^{\alpha}_{\delta}  
\left(
\partial_{\alpha}C_{\beta\gamma}^{\phantom{\beta\gamma}\delta}-
\partial_{\beta}C_{\alpha\gamma}^{\phantom{\alpha\gamma}\delta}
-C_{\alpha\gamma}^{\phantom{\alpha\gamma}\eta}
C_{\beta\eta}^{\phantom{\beta\eta}\delta}+
C_{\beta\gamma}^{\phantom{\beta\gamma}\eta}
C_{\alpha\eta}^{\phantom{\alpha\eta}\delta}.
\right)
\end{split}
\end{flalign}
We will obtain the field equations with the least action principle 
calculating the variations with respect to the metric and both the torsion 
potentials. In order to simplify the variational calculation we proceed 
now in splitting the action in its riemannian part plus torsion-depending 
terms; with the help of (\ref{e:2_4}), we get
\footnote{The symbol ``$\{\}$" stands for riemannian}:
\begin{align}                                        \label{e:3_0}\\
\mathcal{A} =
-\frac{1}{2k}\int dx \ \sqrt{-g}
\left( 
\overset{\{\}}{R}(x)-B^{\alpha\beta\gamma}
B_{\alpha\beta\gamma}-\frac{2}{3}(\partial_{\alpha}\phi)^2
\right);
\end{align}
Variations respect to $g_{\alpha\beta}$, $A_{\alpha\beta}$ and $\phi$ yield:
\begin{align}
\begin{split}
-\overset{\{\}\phantom{^{\alpha\beta}}}{G^{\alpha\beta}}-
\frac{1}{2}g^{\alpha\beta}B^{\mu\nu\sigma}B_{\mu\nu\sigma}&+
3B^{\alpha\nu\sigma}B^{\beta}_{\phantom{\beta}\nu\sigma}+ \\&
-\frac{8}{3}\left(
\frac{1}{2}g^{\alpha\beta}
(\partial_{\mu}\phi)^2-
g^{\alpha\mu}g^{\beta\nu}(\partial_{\mu}\phi)(\partial_{\nu}\phi)
\right) =0\\ 
\end{split}                                     
                                                      \label{e:3_1}\\
&\overset{\{\}}{D}_{\mu}B^{\mu\alpha\beta} 
=0                                                                 
                                                      \label{e:3_2}\\
&\overset{\{\}}{D}_{\mu}g^{\mu\nu}\partial_{\nu}\phi 
=0.                                                                                                                                       
                                                      \label{e:3_3}
\end{align}
In the first equation (\ref{e:3_1}) there is the Riemannian Einstein 
tensor as in GR; moreover in this case we have four terms all quadratic 
in the torsion potentials. By virtue of this if we are interested in 
solving it at first order for little values of the torsion potentials 
we can neglect those quadratic terms and fall back in the GR field 
equations; we can solve them and get the metric to be put in equations 
(\ref{e:3_2}) and (\ref{e:3_3}) to find, at the first order, the torsion 
potentials. As for equations (\ref{e:3_2}) and (\ref{e:3_3}), it can be 
seen that the goal of a propagating torsion has been reached, in fact we 
have two second order PDE for both the potentials. Equation (\ref{e:3_2}) 
can be simplified using its invariance under the gauge transformation:
\begin{align}
A_{\alpha\beta} \to 
A'~_{\alpha\beta}=A_{\alpha\beta}+\overset{\{\}}{D}_{\alpha}Y_{\beta}-
\overset{\{\}}{D}_{\beta}Y_{\alpha};
\end{align}
in fact, if we choose $Y$ such that
\begin{align}
                                                     \label{e:3_4}
\overset{\{\}}{D}_{\alpha}A'~^{\alpha\beta}=0.
\end{align}
after some calculation involving commutation rules for covariant 
derivatives, it can be put in the form: 
\begin{align}
                                                     \label{e:3_5}
\overset{\{\}}{D}_{\mu}
\overset{\{\}\phantom{^{\mu}}}{D^{\mu}}
A'~_{\alpha\beta}-
\overset{\{\}\phantom{^{\mu\alpha\beta\rho}}}
{R_{\mu\alpha\beta}^{\phantom{\mu\nu\beta}\rho}}
A'~_{\rho}^{\phantom{\rho}\mu}+
\overset{\{\}}{R}_{\alpha\rho}A'~_{\beta}^{\phantom{\beta}\rho}+
\overset{\{\}}{R}_{\beta\rho}A'~^{\rho}_{\phantom{\rho}\alpha}-
\overset{\{\}\phantom{^{\mu\beta\alpha\rho}}}
{R_{\mu\beta\alpha}^{\phantom{\mu\beta\alpha}\rho}}
A'~^{\mu}_{\phantom{\mu}\rho}\equiv \Delta_{DR}(A') = 0.
\end{align}
Here $\Delta_{DR}$ is the De Rham operator which generalize the Laplacian 
operator in non flat spaces. It is easy to show that a field $A'$ obeying 
(\ref{e:3_5}) and (\ref{e:3_4}) has only one freedom degree, or, in other words, 
it can have only one direction of polarization in the iperplane normal to its 
propagation direction.
As far as equation (\ref{e:3_3}) is concerned, it is worth noting that it 
is a massless Klein-Gordon field equation so that we can consider the potential 
$\phi$ as a geometrical manifestation of this field.

\section{Test particles paths}                                    \label{s:4}

The problem of the determination of the equation of motion of a test 
particle can be approached in a number of ways; one is that proposed 
by Papapetrou \cite{ppptr} which consists in obtaining the equation 
of motion from the conservation law of the energy-momentum tensor. 
According to us this approach has some unsatisfactory aspects: the 
first is that we can have some ambiguity on the right way to get
the conservation law of the energy-momentum tensor since we can start 
both using N\"{o}ether theorem and  Ricci identities, but, in spaces 
with torsion, the results can be different \cite{hhl3}; in addition, 
once we have the conservation law, we must have the expression of 
the energy-momentum tensor which is rather difficult especially in 
the case of presence of nonriemannian quantities as torsion; in this 
case, in fact, a anti-symmetric part of the energy-momentum tensor, 
probably depending on spin, is involved and it 
is not clear either whether it is correct to give a semiclassical 
expression of it, being spin a purely quantum quantity, or its explicit 
form. Than we have the approach of Hojman \cite{hjmn2} which consists in 
defining all the possible scalar quantities that can be in the test 
particle action; then the action is build up and the equation of motion 
are obtained by variations with respect to the particle coordinates. 
According to us this approach has one of the same unsatisfactory aspects 
of the previous, since, taking into account the spin of the test particle, 
again we need to have a semiclassical expression for the spin depending 
part of the test particle action although spin should be treated only in 
quantum mechanics. Another way to get the test particle equation of 
motion is to make use of the principle of the shortest path which assumes 
that a test particle moves from a point A to another B in space in a way 
such that its trajectory has the least length among all the curves 
joining A with B. 
Although this method seems simple and nice, it is completely regardless 
of the presence of torsion because the last, not contributing to the 
length of a curve, neither appears in the equation of motion of any test 
particle.\\
According to us, instead, the presence of a tensorial quantity as 
torsion, which has a role in the parallel transport of vector fields in 
space, should have some effects on the motion and so we assume as the 
correct method to have the equation of motion in curved space from the 
knowledge of that in flat space the following minimal 
substitution rule:
\begin{align}                                                  \label{e:5_1}
\text{Ordinary derivative} \ \left(\frac{d}{d\tau}\right) \to 
\text{Covariant derivative} \ \left(\frac{D}{d\tau}\right).
\end{align}
According to this rule the equation of motion in curved space is obtained 
from the analogous one of special relativity
\begin{align}
\frac{du^{\alpha}}{d\tau}=0,                                      \label{e:5_2}
\end{align}
where $u^{\alpha}$ is the 4-velocity via the (\ref{e:5_1}):
\begin{align}                   
\frac{Du^{\alpha}}{d\tau}=0
\end{align}
which can be rewritten as:
\begin{align}                                           \label{e:5_3}
\frac{du^{\alpha}}{d\tau}=
-C_{\mu\nu}^{\phantom{\mu\nu}\alpha}u^{\mu}u^{\nu}=
-\Gamma_{\mu\nu}^{\phantom{\mu\nu}\alpha}u^{\mu}u^{\nu}-g^{\alpha\lambda}
\frac{2}{3}((\partial_\lambda\phi)g_{\mu\nu}-
(\partial_\nu\phi)g_{\mu\lambda})u^{\mu}u^{\nu}.
\end{align}
This is the autoparallel equation which, together with the geodesic 
equation, is a special curve that can be defined in non flat spaces; 
while the latter is the shortest curve joining two points, the former 
is the curve whose tangent vector is parallel transported along it. 
The autoparallel curve is the simplest generalization of the flat space 
equation of motion (\ref{e:5_2}) which is suitable to take into account of 
torsion or other nonriemannian quantities
\footnote{In this contest it is obvious how to include nonmetricity.}.
\\
\\
It is worth noting that it is also possible to introduce a new action 
principle such that, starting from the action 
\begin{align}                                                   \label{e:5_4}
\mathcal{A}^M=-\frac{M}{2}\int_{\tau_1}^{\tau_2}d\tau \ \dot{x}^2,
\end{align}
where $\tau$ is the proper time, we can have autoparallels as the right 
trajectories. This approach is proposed in \cite{klnrt1},\cite{klnrt2},
\cite{klnrt3} and here we summarize it briefly. The key point is that a 
spacetime with torsion, which can be obtained from a flat spacetime by 
a nonholonomic mapping, is affected by a closure failure of parallelograms; 
as a consequence, variations of test particle trajectories cannot be 
performed as in the usual way in flat spacetime, i.e. keeping $\delta x^a(\tau)$ 
vanishing at endpoints. 
In fact, \emph{the variations} $\delta^S q^{\mu}(\tau)$, images of $\delta x^a(\tau)$ 
under a nonholonomic mapping, \emph{are generally not closed}; so they can be chosen 
to be zero at the initial point but then they are nonvanishing at the final 
point, this failure being proportional to torsion. The variational 
calculation, then, can be done as follows; first we rewrite explicitly the 
relation between the old ($x^a(\tau)$) and the new ($q^{\mu}(\tau)$) paths 
in integral form:
\begin{align}                                                  \label{e:5_5}
q^{\mu}(\tau)=q^{\mu}(\tau_1)+
\int_{\tau_1}^{\tau} d\tau' e_a^{\phantom{a}\mu}(q(\tau')) \dot{x}^a(\tau'),
\end{align}
where $e_a^{\phantom{a}\mu}(q(\tau'))$ is the nonholonomic mapping; then we 
calculate the variation associated to $q^{\mu}(\tau)$:
\begin{align}
\delta^Sq^{\mu}(\tau)=
\int_{\tau_1}^{\tau} d\tau' 
\left[\left(\delta^S e_a^{\phantom{a}\mu}(q(\tau'))\right) \dot{x}^a(\tau') + 
e_a{}^\mu(q(\tau')) \delta \dot{x}^a(\tau')
\right]
\end{align}
After having introduced a further quantity called auxiliary nonholonomic 
variation:
\begin{align}
\deltabar q^{\mu}(\tau) \equiv e_a^{\phantom{a}\mu}(q(\tau)) \delta x^a(\tau),
\end{align}
which, in contrast to $\delta^S q^{\mu}(\tau)$, vanishes at endpoints and 
forms closed paths in q-space, we derive the relation
\begin{align}
\begin{split}
\frac{d}{d\tau}\delta^S q^{\mu}(\tau)=
\left(\delta^S e_a^{\phantom{a}\mu}(q(\tau))\right) \dot{x}^a(\tau) + 
e_a{}^\mu(q(\tau)) \delta \dot{x}^i(\tau)= \\
=[ \delta^S e_a{}^\mu(q(\tau)) ]\dot{x}^i(\tau )
+ e_a{}^\mu (q(\tau)) \frac{d}{d\tau}[e^a{}_ \nu (q(\tau)) \deltabar
q^ \nu (\tau)],
\end{split}
\end{align}
which, after the substitutions:
\begin{align}
\delta^S  e_a{}^ \mu = -C_{ \lambda  \nu }{}^ \mu \delta^S q^ \lambda
e_a{}^ \nu  
\hspace{1em},
\hspace{1em}
\frac{d}{d\tau  } e^i{}_\nu = C_{ \lambda   \nu  }{}^ \mu  \dot q^ \lambda
e^i{}_\mu,
\end{align}
can be rewritten:
\begin{align}
\frac{d}{d\tau } \delta^S    q^\mu =  -C_{ \lambda  \nu
}{}^ \mu \delta^S q^ \lambda \dot q^ \nu   +
C_{ \lambda    \nu  }{}^ \mu  \dot q^ \lambda  \deltabar q^ \nu
+ \frac{d}{d\tau } \deltabar  q ^\mu  ,
\end{align}
or, better,
\begin{align}                                                  \label{e:5_6}
\frac{d}{d\tau } \delta^S b^ \mu  = -
C_{\lambda \nu }{}^\mu  \delta^S b ^\lambda
\dot{q}^\nu  + 2S _{ \lambda \nu  }{}^\mu
\dot{q}^ \lambda  \deltabar q^ \nu  ,                       
\end{align}
where we have introduced $\delta^S b^{\mu}$, the difference between 
$\delta^S q^ \mu$ and $\deltabar q^ \mu$.
Now we can calculate the variation of the action analogous of (\ref{e:5_4}) 
under a nonholonomic variation $\delta^S q^{\mu} = \deltabar q^{\mu}+
\delta^S b^{\mu}$:
\begin{align}
\delta^S \mathcal{A}^M \equiv
\delta^S \left( 
- \frac{M}{2}\int_{\tau_1}^{\tau_2}  d\tau \  
g_{\mu\nu}\dot{q}^{\mu}\dot{q}^{\nu}
 \right)
= -M\int^{\tau_2}_{\tau_1} d\tau  \left( g_{\mu \nu }
               \dot{q}^\nu \delta^S \dot{q}^\mu + \frac{1}{2}
              \partial _\mu g_{\lambda \kappa }
             \delta^S q^\mu   \,  \dot{q}^\lambda \dot{q}^\kappa \right);
\end{align}
using $[\delta^S,d/d\tau]=0$ which follow from (\ref{e:5_5}), we can partially 
integrate the $\delta^S q-$term and, by the identity $\partial_{\mu} 
g_{\nu\lambda} \equiv C_{\mu\nu\lambda}+C_{\mu\lambda\nu}$, get:
\begin{align}
\delta^S \mathcal{A}^M
=- M \int^{\tau_2}_{\tau_1} d\tau \left[ 
- g_{\mu\nu} \left(
\ddot{q}^\nu  + \Gamma_{\lambda\kappa}{}^\nu
\dot{q}^\lambda \dot{q}^\kappa \right)\deltabar q^ \mu
+\left(g_{\mu\nu} \dot q^{\nu} \frac{d}{d\tau } \delta^S b^{\mu} +
C _{\mu\lambda\kappa}
\delta^S b^{\mu} \dot{q}^ \lambda \dot{q} ^ \kappa
\right)
\right].
\end{align}
Now it is straightforward to obtain first the equation of motion in 
absence of torsion ($\delta^S b^\mu (\tau ) \equiv 0$):
\begin{align}
 \ddot{q}^\nu  +
          \Gamma _{\lambda \kappa }{}^\nu \dot{q}^\lambda
           \dot{q}^\kappa =0,
\end{align}
that is the geodesic equation; then we can consider all the affine 
connection and get, with the help of (\ref{e:5_6}), the autoparallel equation:
\begin{equation}
\ddot{q}^\nu  +
C _{\lambda \kappa }{}^\nu \dot{q}^\lambda
\dot{q}^\kappa =0.
\end{equation}
\\
As for the possibility to get the autoparallel motion from the 
energy-momentum tensor conservation, we now give a possible modification of 
the test particle action such that this result could be partially obtained. 
We assume the test particle action to be:
\begin{align}
\mathcal{A}^{M}=\int d\tau \ u^ \mu u^ \nu \ e^{-\frac{1}{4}\phi} g_{\mu\nu};
\end{align}
then we can calculate its variations with respect to $g_{\mu\nu}$ and $\phi$:
\begin{align}
\begin{split}                                              \label{e:5_7}
{}^{g}T^{\mu\nu} = 
\frac{\delta \mathcal{A}^{M}}{\delta g_{\mu\nu}}=
\int d\tau \ \frac{u^ \mu u^ \nu}{\sqrt{-g}} \ 
e^{-\frac{1}{4}\phi} \ \delta(x-x_0)       ,       \\
{}^{\phi}T=
\frac{\delta \mathcal{A}^{M}}{\delta \phi}=
-\frac{1}{4} \int d\tau \ \frac{u^ \mu u^ \nu}{\sqrt{-g}} \ 
e^{-\frac{1}{4}\phi} \ g_{\mu\nu} \ \delta(x-x_0)  ,    
\end{split}
\end{align}
from comparison with
\begin{align}
\delta {\cal A} = \int d^4 x \sqrt{-g} \ 
({}^{g}T^{\mu\nu} \ \delta g_{\mu\nu} + {}^{\phi}T \ \delta \phi).
\end{align}
Now, following the same procedure of Hammond \cite{hmmnd}, we consider the 
motion of a test particle in a background geometry which is perturbed in 
a negligible way by it and start from the identity:
\begin{align}
(\sqrt{-g} \ {}^{g}T^{\mu\nu})_{,\nu}=
\sqrt{-g} \ {}^{g}T^{\mu\nu}_{\phantom{\mu\nu};\nu}-
\sqrt{-g}\Gamma_{\alpha\beta}^{\phantom{\alpha\beta}\mu}
 \ {}^{g}T^{\alpha\beta}; 
\end{align}  
now we integrate it over a small volume $dV$ where the only appreciable 
energy momentum tensor is that of the test particle and, discarding all 
surface terms, with the help of the conservation law in our case:
\begin{align}
{}^{g}T^{\mu\nu}_{\phantom{\mu\nu};\nu}=
\frac{8}{3}\partial^{\mu}\phi \ {}^{\phi}T,
\end{align}
we get:
\begin{align}
\frac{1}{u^0}\frac{d}{d\tau}\int dV(\sqrt{-g} \ {}^{g}T^{(\mu 0)})=
\frac{8}{3}\partial^{\mu}\phi \int dV(\sqrt{-g} \ {}^{\phi}T)-
\Gamma_{\alpha\beta}^{\phantom{\alpha\beta}\mu}
\int dV(\sqrt{-g} \ {}^{g}T^{(\alpha\beta)}).
\end{align}
This, with the help of (\ref{e:5_7}), can be rewritten in the form:
\begin{align}
                                                                  \label{e:5_8}
\frac{du^{\alpha}}{d\tau}=
-\Gamma_{\mu\nu}^{\phantom{\mu\nu}\alpha}u^{\mu}u^{\nu}-
\frac{2}{3} g^{\alpha\lambda}(\partial_\lambda\phi)g_{\mu\nu}u^{\mu}u^{\nu}.
\end{align}
If we multiply both members of this equation by $u_{\alpha}$ we get:
\begin{align}
                                                                 \label{e:5_9}
0=u_{\alpha}\partial^{\alpha}\phi.
\end{align}
Now if we consider our autoparallel equation (\ref{e:5_3}) we see that it 
is in accordance with these last two equations (\ref{e:5_8}),(\ref{e:5_9}), 
meaning that the Papapetrou motion is included as a special case.

\section
{Autoparallels, Autoparallel deviation and their nonrelativistic limit}
                                                                   
                                                                   \label{s:5}

At the end of the last section we have concluded, on the basis of our 
minimal substitution rule, that test particles follow autoparallel 
trajectories whose equation we recall now:
\begin{align}
\frac{d^2x^{\alpha}}{d\tau^2}=
-\Gamma_{\beta\gamma}^{\phantom{\beta\gamma}\alpha}
\frac{dx^{\beta}}{d\tau}\frac{dx^{\gamma}}{d\tau}
-K_{\beta\gamma}^{\phantom{\beta\gamma}\alpha}
\frac{dx^{\beta}}{d\tau}\frac{dx^{\gamma}}{d\tau}.           \label{e:6_1}
\end{align}
It is easy to see that in this expression the antisymmetric part of 
torsion contribution vanishes; it only contributes as a source for the metric 
through (\ref{e:3_1}).
In this section we will study the nonrelativistic limit of autoparallels 
and in addition we will calculate the analogous of the geodesic deviation 
and we will see the role of torsion in the tidal forces.

\subsection[Nonrelativistic limit of autoparallels]
{Nonrelativistic limit of autoparallels}
In order to calculate the nonrelativistic limit of (\ref{e:6_1}), we make 
the following assumptions: 
\begin{itemize}
\item[a)]3-velocity much smaller than c so that we can put $u^i\simeq v^i$;
\item[b)]the gravitational field is static and weak, i. e. the quantities 
$h_{\alpha\beta}=g_{\alpha\beta}-\eta_{\alpha\beta}$ (here 
$\eta_{\alpha\beta}$ 
is the Minkovsky metric) are very small;
\item[c)]the torsion potential $\phi$ is static and weak.
\end{itemize}
By virtue of these assumptions, since we want to keep only first order 
terms, we will eliminate all those terms containing $\phi$, $h_{\alpha\beta}$ 
e $v^i$ more than once. After some calculations we obtain:
\begin{align}                                               \label{e:6.1_1}
\frac{d^2x_{i}}{dt}=-
\frac{\kappa}{2}\partial_{i}h_{00}-
\frac{2}{3}\partial_{i}\phi.
\end{align}
Now we recall that in GR we had:
\begin{align}       
\frac{d^2x_{i}}{dt}=-\frac{\kappa}{2}\partial_{i}h_{00},
\end{align}
that allowed us to identify $h_{00}$ with the gravitational potential 
$\Phi$:
\begin{align}
\frac{\kappa}{2}h_{00}=\Phi.
\end{align}
As we can see from (\ref{e:6.1_2}) the ``force" due to the torsion 
potential is present in the same form of the gravitational field $h_{00}$;  
in addition, as for the order we are interested in, and reminding of the 
supposed field's static nature, equation (\ref{e:3_3}) for the field $\phi$ 
reduces to:
\begin{align}                                   \label{e:6.1_2}
\Delta\phi(\vec{x})=0, 
\end{align}
which is the same of the gravitational field:
\begin{align}                                   \label{e:6.1_3}
\Delta h_{00}(\vec{x})=4\pi\rho.
\end{align}

\subsection[Deviation of autoparallels]{Deviation of autoparallels}
Following standards calculations \cite{rffn} and reminding that now test 
particles move along autoparallels, if we take two of them initially very 
close each other, we find their relative acceleration to be:   
\begin{align}
\frac{\overset{\{\}\phantom{^2}}{D^2}s^{\alpha}}{d\tau^2}=
&-\overset{\{\}\phantom{^{\delta\gamma\beta\alpha}}}
{R_{\delta\gamma\beta}^{\phantom{\delta\gamma\beta}\alpha}}
s^{\delta}u^{\beta}u^{\gamma}+\\
&-K_{\beta\gamma}^{\phantom{\beta\gamma}\alpha}\left(
\frac{ds^{\gamma}}{d\tau}u^{\beta}+
\frac{ds^{\beta}}{d\tau}u^{\gamma}\right)-
\left(
\overset{\{\}}{D}_{\delta}K_{\beta\gamma}^{\phantom{\beta\gamma}\alpha}\right)
s^{\delta}u^{\beta}u^{\gamma}.
\end{align}
Here $s^{\alpha}$ is an infinitesimal vector representing the relative 
displacement between the two particles.
Now we can substitute our expression for the contortion tensor (\ref{e:2_3}) 
and get:
\begin{align}
\begin{split}                                                
                                                             \label{e:6.2_1}
\frac{\overset{\{\}\phantom{^2}}{D^2}s^{\alpha}}{d\tau^2}=
-\overset{\{\}\phantom{^{\delta\gamma\beta\alpha}}}
{R_{\delta\gamma\beta}^{\phantom{\delta\gamma\beta}\alpha}}
s^{\delta}u^{\beta}u^{\gamma}+\\
-\frac{2}{3}\left[
\delta_{\beta}^{\phantom{\beta}\alpha}
\left(\partial_{\gamma}\phi\right)+
g^{\alpha\eta}g_{\beta\gamma}
\left(\partial_{\eta}\phi\right)\right]
\left(
\frac{ds^{\gamma}}{d\tau}u^{\beta}+
\frac{ds^{\beta}}{d\tau}u^{\gamma}\right)+\\
-\frac{2}{3}
\left[
\delta_{\beta}^{\phantom{\beta}\alpha}
\overset{\{\}}{D}_{\delta}\left(\partial_{\gamma}\phi\right)+
g^{\alpha\eta}g_{\beta\gamma}\overset{\{\}}{D}_{\delta}
\left(\partial_{\eta}\phi\right)\right]
s^{\delta}u^{\beta}u^{\gamma}.
\end{split}
\end{align}
This equation represents the generalization to a theory with torsion of 
the geodesic deviation of GR:
\begin{align}
\frac{\overset{\{\}\phantom{^2}}{D^2}s^{\alpha}}{d\tau^2}=
-\overset{\{\}\phantom{^{\delta\gamma\beta\alpha}}}
{R_{\delta\gamma\beta}^{\phantom{\delta\gamma\beta}\alpha}}
s^{\delta}u^{\beta}u^{\gamma}.
\end{align}
It is easy to see that the difference between the two expressions consists 
in two linear terms in torsion and torsion derivative, the first of which 
is multiplied by a term of relative velocity.
Once again we see completely antisymmetric part of torsion not 
contributing, if not as a source in the field equation (\ref{e:3_1}), to 
this expression.\\

\vspace{1em}
To calculate the nonrelativistic limit of (\ref{e:6.2_1}), we will make the 
same assumptions b), c) made in the last paragraph plus:
\begin{itemize}
\item[a$\grave{}$)]involved velocities are much smaller than c and:
\begin{align*}
\frac{dx^{\alpha}}{d\tau}\simeq(1,0,0,0);
\end{align*}
\item[d)] $s^0=\frac{ds^0}{d\tau}=0$, which simply means that the 
particles accelerations are compared at the same time.
\end{itemize}
In this case, by virtue of our assumptions, we will keep only terms 
containing at most, as factors multiplied by $s^i$, one between 
$h_{\alpha\beta}$ and $\phi$. Then, (\ref{e:6.2_1}) reduces to:
\begin{align}
\frac{d^2s^i}{dt^2}\simeq
-\overset{\{\}\phantom{^{j00i}}}
{R_{j00}^{\phantom{\delta00}i}}s^{j}-
\frac{2}{3}\eta^{ij}\left(\partial_{kj}\phi\right)s^k;
\end{align}
so the tidal field is:
\begin{align}
\mathcal{G}^i=-\overset{\{\}\phantom{^{j00i}}}
{R_{j00}^{\phantom{j00}i}}s^{j}-
\frac{2}{3}\eta^{ij}\left(\partial_{k}\partial_{j}\phi\right)s^k.
\end{align}
Now, from GR, it is known that in the nonrelativistic limit:
\begin{align}
\overset{\{\}\phantom{^{j00i}}}{R_{j00i}}=\partial_{j}\partial_{i}\Phi,
\end{align}
where $\Phi$ is the gravitational potential.\\
Taking that into account we can rewrite the final expression for the tidal 
field:
\begin{align}                                              \label{e:6.2_2}
\mathcal{G}_i=-\left(\partial_{i}\partial_{j}\Phi\right)s_j-
\frac{2}{3}\left(\partial_{i}\partial_{j}\phi\right)s_j.
\end{align}
So, in the nonrelativistic limit, torsion produces a tidal forces 
effect analogous to the one produced by the gravitational field.\\

It must be noted, now, that since the fields $h_{00}$ and $\phi$, in the 
nonrelativistic limit, both obey a Poisson PDE ((\ref{e:6.1_2}) and 
(\ref{e:6.1_3})) and enter in equations (\ref{e:6.1_1}) and (\ref{e:6.2_2}) 
in the same way, it is impossible to distinguish the effect of the 
torsion field from that of the gravitational one unless we know exactly 
the source and the initial condition for the latter; this, together with 
the small intensity of the torsion forces, makes them even more difficult 
to be detected.

\section{Concluding remarks}
We now want to summarize briefly the main results stated here. First we 
have exposed a formulation of a geometric theory of the GR type in the vacuum 
which is able to predict propagating torsion; 
than, in order to determine the equation of motion of test particle in presence 
of this new geometric quantity, we have established a principle of minimal 
substitution (\ref{e:5_1}) which implied autoparallels were the right 
trajectories. 
Finally we have determined the analogous of the geodesic equation for the 
autoparallels and calculated the nonrelativistic limit of both this deviation 
(\ref{e:6.2_2}) and the autoparallels (\ref{e:6.1_1}) showing that in those 
expressions, and also in the field equation (\ref{e:6.1_2}), in this limit, 
the field $\phi$ enters in the same way as the gravitational field $h_{00}$ 
making itself difficult to be detectable.

}


\begin{thebibliography}{99}

\bibitem{bn}
   Aubin, T., \emph{A course in differential geometry}, Graduates studies 
   in mathematics, vol.~27, 
    American Mathematical Society, 2000.


\bibitem{blgjvc}
   Blagojevic, M., \emph{Gravitation and Gauge Symmetries}, Institute of 
   Physics Publishing, Bristol, UK (2002).


\bibitem{crtn1}
   Cartan, \'E, \emph{Sur une g\'en\'eralization de la notion de courbure 
de Riemann et les espaces \`a torsion}, C. R. Acad. Sci. (Paris)  
\textbf{174}, 
    (1922), 593. 


\bibitem{crtn2}
   Cartan, \'E, \emph{Sur les vari\'et\'es \`a connexion affine et la 
th\'eorie de la relativit\'ee g\'eneralis\'ee \textrm{I}, \textrm{I} 
(suite)}, Ann. Ec. Norm. Sup. \textbf{40} 
    (1923) 325; \textbf{41} (1924), 1.


\bibitem{crtn3}
   Cartan, \'E, \emph{Sur les vari\'et\'es \`a connexion affine et la 
th\'eorie de la relativit\'ee g\'eneralis\'ee \textrm{II}}, Ann. Ec. Norm. 
Sup. \textbf{42}
     (1925), 17.


\bibitem{ddngtn}
   Eddington, A., S., \emph{A generalization of Weyl's theory of 
electromagnetic and gravitational fields}, Proc. R. Soc. Lond. 
\textbf{A99}
    (1921), 104.


\bibitem{nstn}
   Einstein, A., \emph{The meaning of Relativity}, (Princeton University 
   Press, Princeton, NY), 5th Ed., 1956.


\bibitem{hhl1}
   Hehl, F., W., von der Heide, P., Kerlick, G., D., Nester, J., M., 
\emph{General 
   Relativity with spin and torsion: Foundations and prospects}, Rev. Mod. 
   Phys. \textbf{48} 
    (1976), 393.


\bibitem{hhl2}
   Hehl, F., W., McCrea, J., D., Mielke, E., W., Ne'eman, Y., 
   \emph{Metric-affine gauge theory of gravity: Field equations, Noether 
   identities, world spinors, and breaking of dilation invariance}, Phys. 
   Rep. \textbf{A192} 
    (1994), 122.


\bibitem{hhl3}
   Hehl, F., W., McCrea, J., D., \emph{Bianchi identities and the 
   automatic conservation of energy-momentum and angular momentum in 
   general-relativistic field theories}, Found. Phys. \textbf{16} 
    (1986), 267.


\bibitem{hmmnd}
   Hammond, R.T., \emph{Spin, Torsion, Forces}, Gen. Rel. Grav. 
\textbf{26} 
    (1994), 247.

\bibitem{hjmn}
   Hojman, S., Rosenbaum, M., Ryan, M., P., \emph{Propagating torsion and 
   gravitation},  Phys. Rev. D
\textbf{19} 
    (1979), 430.
 
\bibitem{hjmn2}
   Hojman, S., \emph{Lagrangian theory of the motion of spinning particles 
   in torsion gravitational theories},  Phys. Rev. D
\textbf{18} 
    (1978), 2741.

\bibitem{kbbl}
   Kibble, T.W.B., \emph{Lorentz invariance  and the gravitational 
field}
   J. Math. Phys. \textbf{2} 
    (1961), 212.

\bibitem{klnrt1}
   Kleinert, H., Fiziev, P., \emph{New action principle for 
   classical particle trajectories in spaces with torsion}, 
   Europhys. Lett. {\bf 35},
   241 (1996) (hep-th/9503074).


\bibitem{klnrt2}
   Kleinert, H., Pelster, A., \emph{Autoparallels from a new action 
   principle}, 
   Gen. Rel. Grav. \textbf{31} 
    (1999), 1439 (gr-qc/9605028).

\bibitem{klnrt3}
   Kleinert, H., \emph{Nonholonomic Mapping Principle
for Classical and Quantum Mechanics
in Spaces with Curvature and Torsion}, 
   (gr-qc/0203029).


\bibitem{khfss}
   Kuhfuss, R., Nitsch, J., \emph{Propagating modes in gauge field 
   theories of gravity}, Gen. Rel. Grav. \textbf{18} (1964), 1207.


\bibitem{msnr}
  Misner, C.W., Thorne, K.S., Wheeler, J.A., \emph{Gravitation},   
(Freeman, San Francisco), 
   1973.


\bibitem{nvll}
  Neville, D.E., \emph{Spin-2 propagating torsion}, Phys. Rev. D \textbf{23} 
    (1981), 1244.


\bibitem{ppptr}
   Papapetrou, A., \emph{Spinning test-particles in general relativity}, 
   Proc. Roy. Soc. Lond. \textbf{A209} 
    (1948), 248.

\bibitem{rffn}
   Ohanian, H., C., Ruffini, R., \emph{Gravitazione e Spazio-Tempo}, 
   (Zanichelli, Bologna) 
    (1997), 291.


\bibitem{schtn}
   Schouten, J.A., \emph{Ricci calculus}, (Springer, Berlin), 2nd ed., 
   1954.


\bibitem{schrdngr}
   Schr\"odinger, E., Proc. R. Irish Acad. \textbf{A49} 
    (1943), 43, 135.


\bibitem{scm1}
    Sciama, D.W., \emph{On the analogy between charge and spin in general 
relativity}, Recent Developments in General Relativity, (Pergamon+PWD, 
Oxford), 
    1962, 415.


\bibitem{scm2}
   Sciama, D.W., \emph{The physical structure of general relativity}, Rev. 
Mod. Phys. \textbf{36} 
    (1964), 463 and 1103.


\bibitem{szgn1}
   Sezgin, E., van Nieuwenhuizen, P. \emph{New ghost free gravity 
   lagrangians with propagating torsion}, Phys. Rev. D \textbf{21} (1980), 
   3269.


\bibitem{szgn2}
   Sezgin, E., \emph{Class of ghost free gravity lagrangians with massive 
   or massless propagating torsion}, Phys. Rev. D \textbf{24} (1981), 1677.



\bibitem{tsytln}
   Tseytlin, A.A., \emph{On the Poincare and the de Sitter gauge theories of gravity 
   with propagating torsion}, Phys. Rev. D \textbf{26} (1982), 3327.


\bibitem{tym}
   Utiyama, R., \emph{Invariant theoretical interpretation of interaction}, 
Phys. Rev. \textbf{101} 
    (1956), 1597.


\bibitem{yng}
   Yang, C., N., Mills, R., L., \emph{Conservation of isotopic spin and 
   isotopic gauge invariance}, Phys. Rev. \textbf{96} 
    (1954), 191.


\end{thebibliography}
\end{document}